# Filament propagation length of femtosecond pulses with different transverse modes

N. Kaya,[1,*] M. Sayrac,[1] G. Kaya,[1] J. Strohaber,[1,2] A. A. Kolomenskii,[1] and H. A. Schuessler[1]

[1] Department of Physics, Texas A&M University, College Station, Texas 77843-4242, USA
[2] Department of Physics, Florida A&M University, Tallahassee, Florida 32307, USA
*Corresponding author: necati@physics.tamu.edu

We experimentally studied intense femtosecond pulse filamentation and propagation in water for Gaussian, Laguerre-Gaussian, and Bessel-Gaussian incident beams. These different transverse modes for incident laser pulses were created from an initial Gaussian beam by using a computer generated hologram technique. We found that the length of the filament induced by the Bessel-Gaussian incident beam was longer than that for the other transverse modes under the conditions of the same peak intensity, pulse duration, and the size of the central part of the beam. To better understand the Bessel-Gaussian beam propagation, we performed a more detailed study of the filament length as a function of the number of radial modal lobes. The length increased with the number of lobes, implying that the radial modal lobes serve as an energy reservoir for the filament formed by the central intensity peak.

Light filaments formed by femtosecond laser radiation propagating in nonlinear media [1, 2] facilitate a number of applications, including remote sensing [3-5], attosecond physics [6-8], and lightning control [9]. In such settings, extended filaments are desirable, but although various techniques aiming to prolong their length have been explored [10-12], the substantial extension of optical filaments continues to attract considerable interest, and much still remains to be understood [13].

Since filamentation is a result of the prevailing of the Kerr self-focusing of an intense pulse over the defocusing by the self-generated weak plasma and the effect of free electrons [14], one can expect that different incident laser transverse modes should also exhibit different filament propagation dynamics. The ideal Bessel beams are known to be diffraction-free beams when they propagate in vacuum [15, 16]. Although ideal Bessel beams do not exist, the use of approximate or quasi-Bessel beams has long been suggested in diverse areas of optical physics, since such beams maintain long propagation lengths in optical media by virtue of the strongly suppressed diffraction of their central lobe over long distances [17, 18].

When a Bessel beam is compared to a Gaussian beam with the same beam diameter, it shows a remarkable resistance to diffraction during propagation [19]. Recent theoretical studies on filamentation dynamics in Ar gas of intense femtosecond beams with different transverse modes have shown that the cross-sectional profile of the laser beam in Bessel modes remain undistorted [20] during propagation over long distances, the outer part of the Bessel beam serves as an energy reservoir for the filament that is formed around the central portion [21]. Recently, Scheller and colleagues have experimentally shown that the propagation of a femtosecond laser filament in air can be substantially extended by an appropriate use of a surrounding auxiliary dressing beam, continuously supplying energy to the filament [22]. It is a seminal idea that the length of the energy transmission by the filament can be strongly affected by the transverse profile of a beam. In liquids the Kerr nonlinearity is about two orders of magnitude larger than in gases [23], and therefore the nonlinear effects develop on a shorter distance and require less power. Therefore, in this study we experimentally investigate filament propagation dynamics in water with intense femtosecond pulses of Gaussian beam (GB), Laguerre-Gaussian beam (LGB), and Bessel-Gaussian beam (BGB) profiles with similar peak intensities, pulse durations, and beam diameters. Particular emphasis is placed on the incident BGB and a more detailed study of the mode structure on its propagation.

The transverse modes of GB and LGB and BGB are described respectively by the following spatial field amplitude:

$$E_{GB}(r, z=0, t=0) = E_0 e^{-r^2/w_0^2} \; , \tag{1}$$

$$E_{LGB}(r, z=0, t=0) = E_0 \left[1 - 2(\beta r)^2\right] e^{-(\beta r)^2} \; , \tag{2}$$

$$E_{BGB}(r, z=0, t=0) = E_0 \, J_0(\alpha r) e^{-(\gamma r)^2} \; , \tag{3}$$

where $E_0$ is the peak amplitude, $w_0$ is the radius of the beam profile at the $1/e^2$ level of intensity, and $J_0$ is the zero order Bessel function. The constants $\alpha$, $\beta$ and $\gamma$ are chosen in such a way as to make the FWHM of the central lobes of the different transverse modes equal (for the GB $w_{FWHM} = w_0 \left(2 ln|2|\right)^{1/2}$).

Laser modes were created from an initial GB of a Ti:sapphire laser system (pulse duration of 50 fs, central wavelength of 800 nm, and an output energy of 1 mJ per pulse at a 1 kHz repetition rate) by using a beam expander and computer-generated holograms [24, 25] displayed on a liquid crystal spatial light modulator (Hamamatsu LCOS-SLM X10468-2). The SLM had a resolution of 800x600 pixels (16 mm x 12 mm), and a maximum reflectivity of >95% for radiation between 750 nm and 850 nm. Figure 1 shows an illustration of the experimental setup used. The laser beam passing through a beam expander (not shown) illuminates the SLM with a phase-amplitude encoded hologram set by a computer to produce a desired optical beam in the 1st diffraction order. Such grayscale computer-generated holograms for GB, LGB and BGB, prepared with a MATLAB code, were displayed on the SLM's LCD via a digital visual interface connection. An illustration of the computer-generated hologram used to produce a BGB in the first diffraction order is shown in the inset of Fig. 1.

We note that the SLM creates a quasi-Bessel amplitude-phase distribution that together with the Gaussian distribution of the incident beam produces a BGB, which has a limited cross section and a finite number of lobes. With all optical losses, including the loss on the SLM, which employs off-axis holography to generate the beam modes, the incident power of the Gaussian beam mode with a 300 μm diameter at FWHM on the water cell was measured as 270 mW at the entrance of the cell. Consequently, for the repetition rate of our laser system (1 kHz) and the pulse duration (50 fs), we obtained an input peak power of 5.4 GW per pulse.

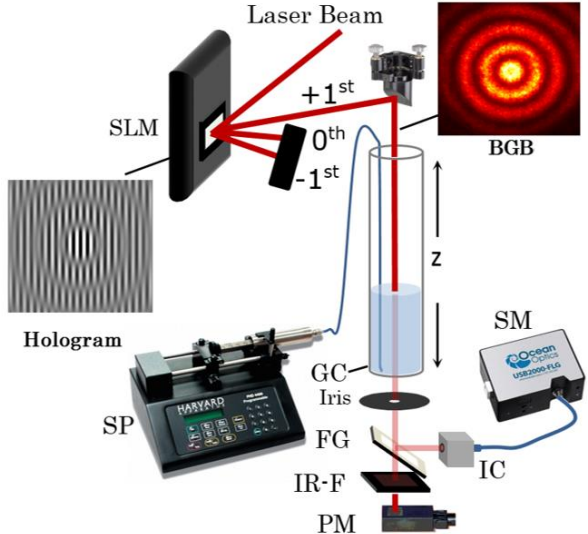

Fig. 1 (Color online) Experimental setup: SLM, spatial light modulator; SP, syringe pump; GC, glass cell with an optical window; FG, flat glass (4% reflectivity at 45 deg.); IR-F or ND-F, infrared filter or neutral density filter to measure the incident beam; PM, power meter; IC, integrated cavity; SM, spectrometer. The inset under the SLM is an example hologram used to create a BGB with an intensity distribution illustrated in the top-right inset.

When the initial power in the produced optical beam exceeds a critical value inside the optical medium, nonlinear optical self-focusing effects become important and initiate a catastrophic collapse of the beam, which is restrained by the plasma formation, inducing defocusing. The filamentation phenomena occur with the resulting dynamic balance of self-focusing and plasma defocusing. For this reason, the sufficient peak power of $P_{in}$ =5.4 GW, which is much larger than the critical power for self-focusing in water, $P_{cr}$=3.6 MW, assured filament formation in water, as was observed previously [26, 27]. Indeed, hot spots in the beam producing white light corresponding to multiple filament formation were directly observed in our experiments. Much care was also taken to ensure that the produced beams had similar peak intensities, pulse durations, and beam diameters. The peak intensity and diameter were determined from the measurements with a power meter interchanged with a CCD camera; the measurements with the latter required additional neutral-density filters in front of it. We kept the distance between the SLM and the glass cell as short as possible to minimize the diffraction of the beams [28]. The beams were passed through a 13 cm-long glass cell, which was arranged vertically to allow us to measure the power and spectrum as a function of the propagation distance by changing the water level in the cell with a programmable infuse/withdraw syringe pump (Harvard PHD 2000). A mechanical iris was used to select the central portion of the beam. An optical flat was positioned after the cell to reflect a small portion of the beam into the integration cavity. Spectral measurements were taken as a function of the propagation distance by collecting the radiation in the cavity with an Ocean Optics USB-2000 spectrometer. Simultaneously, power measurements were obtained using a photodiode power meter head (Ophir PD300-IR) with a spectral range within 700-1800 nm. In order to measure the incident beam power after the water cell, a long-pass glass filter (RG-780) was placed in front of the power meter to filter out white-light. A LabVIEW code was used to control the infusing and withdrawing of water via a syringe pump, and to acquire the values from the power meter and the spectrum. Figure 2 shows spectra of the GB, LGB, and BGB, which were obtained by measuring over the central lobe of the beams as a function of the propagation distance. The corresponding experimentally created beam modes are shown in the right panel of Fig. 2.

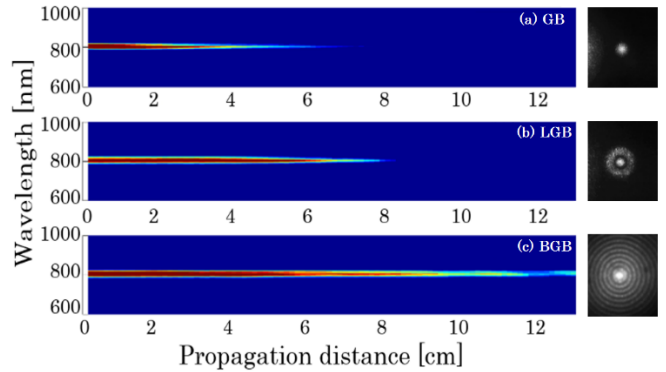

Fig. 2 (Color online) Spectral profiles of the (a) GB, (b) LGB, and (c) BGB measured over the central lobes as a function of the propagation distance. The corresponding experimentally created input beam modes are shown in the right panel.

The self-focusing effect took place near the entrance of the beam into the cell (the self-focusing length of ~ 2cm), and the changes due to nonlinear effects in the spectral profiles of all three types of beams were observed during subsequent propagation. The spectrum for the incident GB mode has the greatest depletion as the pulse propagates in the nonlinear medium followed by that of the incident LGB mode. The spectral profile of the incident BGB exhibits the least depletion among the three transverse modes as a function of distance.

To better understand the propagation of a BGB, we investigated their propagation dynamics by varying the number of radial lobes. Figure 3 depicts the experimentally created beams. The desired number of lobes of BGB was determined by using a radial holographic knife-edge realized with a computer generated hologram on SLM [25]. The power and spectrum were measured from only the central lobe of the resulting BGB.

The central part of the beam had a diameter of 300μm and was similar to the FWHM of the GB. This portion of the beam was selected by the mechanical iris positioned after the water cell and in front of the power meter or the opening of the integration cavity. The spectra measured for BGB as functions of propagation distance are shown in Fig. 4. The effect of increasing number of radial lobes can be seen by an increase in the propagation distance of the central part of the beam. This increase is most noticeable when 1, 2 or 3 lobes are added.

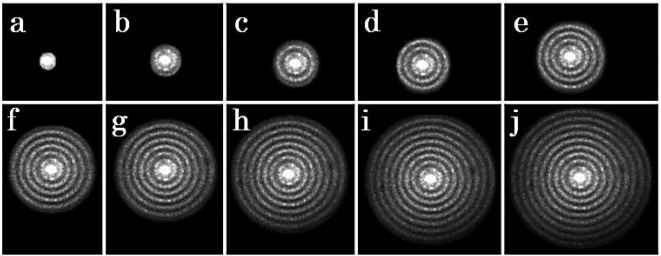

Fig.4. Experimentally created modes of the incident BGB: (a) central peak of the beam with no lobes and (b-j) central peak with additional radial lobes.

In Fig. 5, we show the infrared power measured for only the central peak of the BGB with different number of radial lobes as a function of the propagation distance. With each additional radial lobe we observe the trend of an increasing power delivered to a given propagation distance within the range of distances 2-6 cm. At distances 6-8 cm the delivered power for beams with small number of lobes (1-2) decreases faster, while the beams with multiple lobes (>2) show similar power decay rate, but maintain a higher power, as compared to the beam with just a single central peak.

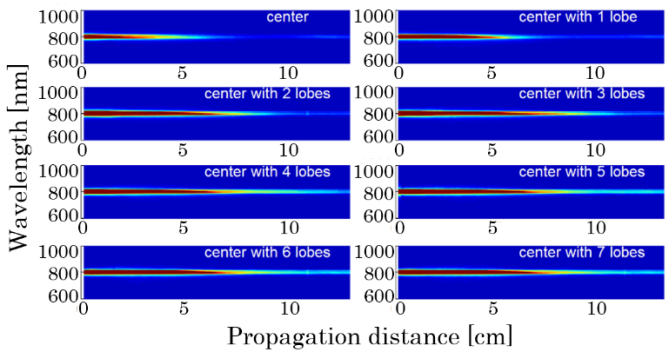

Fig. 5 (Color online) Spectra of the central lobe of the BGB with different number of radial lobes and measured as a function of the propagation distance.

For distances longer than 10cm the delivered power from the beams with 1 and 2 additional lobes (Fig. 3 (b,c)) is close to the power of the beam with no lobes (Fig. 3 (a)), while the power of the beams with multiple lobes (>3) preserves a level higher than that for the beam with no lobes.

The trend of propagation distance elongation was recently demonstrated experimentally by using dressed beams, where the central Gaussian beam is surrounded by auxiliary dressing beam, which is wider and has a lower intensity [22]. Since a Bessel-like beam, which has an annular structure, possesses an inward energy flux towards its optical axis [29], it is expected to be well suited to replenish the filament core, as is confirmed by our measurements. Also, when we compare our results with recent theoretical studies on filamentation of femtosecond beams with different transverse modes in Ar gas [20, 21], we see a similar effect that the central core in BGB mode is sustained for a longer propagation distance (compared to GB and LGB). In addition, we have shown that increasing the outer part of the BGB helps to maintain the energy in the central peak, thus this outer part serves as an energy reservoir for the filaments formed in the central portion of the beam.

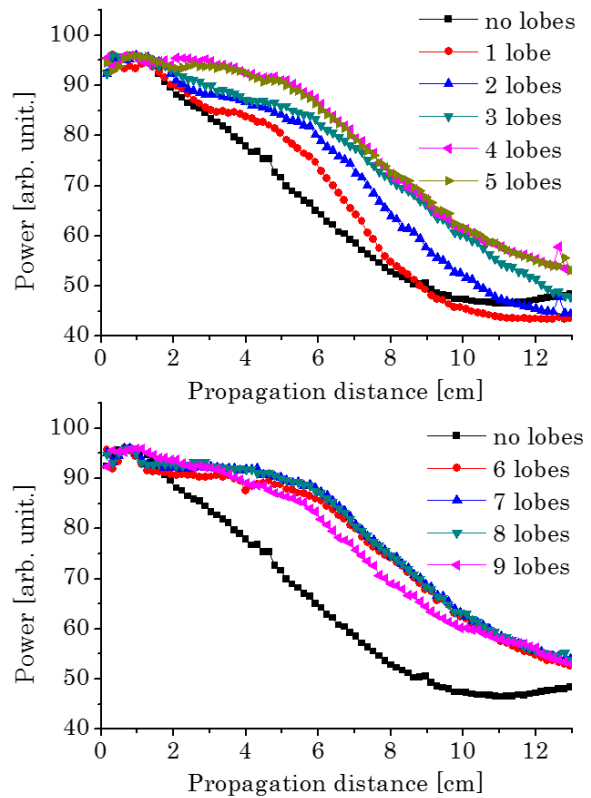

Fig. 3 (Color online) IR power measured for the central peak of the BGB with the different number of radial lobes as a function of the propagation distance.

In conclusion, we experimentally investigated the propagation of intense femtosecond pulses in water for different incident beam profile configurations. Among the three different transverse modes (GB, LGB and BGB), the length of propagation of BGB for similar characteristics of the central part of the beam (peak intensity, pulse duration, and beam diameter) is considerably longer than for the other beams as determined by the measurements of the decay of the spectral and total power. A more detailed study of the BGB with different number of the radial lobes indicates that the outer part of the BGB serves as an energy reservoir, so that the filaments formed at the inner part can persist for a long propagation distance with increasing number of radial lobes. Our findings clearly indicate the high potential of BGB for various nonlinear optics applications involving the propagation of ultrafast pulses in a Kerr medium.

This work was supported by the Robert A. Welch Foundation Grant No. A1546 and the Qatar Foundation under the grant NPRP 6 - 465 - 1 - 091.

References

1. A. Couairon and A. Mysyrowicz, "Femtosecond filamentation in transparent media," Phys Rep **441**, 47-189 (2007).
2. M. Mlejnek, E. M. Wright, and J. V. Moloney, "Dynamic spatial replenishment of femtosecond pulses propagating in air," Opt. Lett. **23**, 382-384 (1998).
3. Q. Luo, H. L. Xu, S. A. Hosseini, J. F. Daigle, F. Théberge, M. Sharifi, and S. L. Chin, "Remote sensing of pollutants using femtosecond laser pulse fluorescence spectroscopy," Appl. Phys. B **82**, 105-109 (2006).
4. H. L. Xu and S. L. Chin, "Femtosecond laser filamentation for atmospheric sensing," Sensors **11**, 32-53 (2011).
5. J. Kasparian, M. Rodriguez, G. Méjean, J. Yu, E. Salmon, H. Wille, R. Bourayou, S. Frey, Y.-B. André, A. Mysyrowicz, R. Sauerbrey, J.-P. Wolf, and L. Wöste, "White-Light Filaments for Atmospheric Analysis," Science **301**, 61-64 (2003).
6. F. Krausz and M. Ivanov, "Attosecond physics," Reviews of Modern Physics **81**, 163-234 (2009).
7. A. Pierre and F. D. Louis, "The physics of attosecond light pulses," Reports on Progress in Physics **67**, 813 (2004).
8. E. Goulielmakis, M. Schultze, M. Hofstetter, V. S. Yakovlev, J. Gagnon, M. Uiberacker, A. L. Aquila, E. M. Gullikson, D. T. Attwood, R. Kienberger, F. Krausz, and U. Kleineberg, "Single-Cycle Nonlinear Optics," Science **320**, 1614-1617 (2008).
9. R. Ackermann, G. Méchain, G. Méjean, R. Bourayou, M. Rodriguez, K. Stelmaszczyk, J. Kasparian, J. Yu, E. Salmon, S. Tzortzakis, Y. B. André, J. F. Bourrillon, L. Tamin, J. P. Cascelli, C. Campo, C. Davoise, A. Mysyrowicz, R. Sauerbrey, L. Wöste, and J. P. Wolf, "Influence of negative leader propagation on the triggering and guiding of high voltage discharges by laser filaments," Appl. Phys. B **82**, 561-566 (2006).
10. P. Polynkin, M. Kolesik, A. Roberts, D. Faccio, P. Di Trapani, and J. Moloney, "Generation of extended plasma channels in air using femtosecond Bessel beams," Optics express **16**, 15733-15740 (2008).
11. O. G. Kosareva, A. V. Grigor'evskii, and V. P. Kandidov, "Formation of extended plasma channels in a condensed medium upon axicon focusing of a femtosecond laser pulse," Quantum Electronics **35**, 1013 (2005).
12. S. Akturk, B. Zhou, M. Franco, A. Couairon, and A. Mysyrowicz, "Generation of long plasma channels in air by focusing ultrashort laser pulses with an axicon," Optics Communications **282**, 129-134 (2009).
13. A. Schweinsberg, J. Kuper, and R. W. Boyd, "Loss of spatial coherence and limiting of focal plane intensity by small-scale laser-beam filamentation," Physical Review A **84**, 053837 (2011).
14. S. L. Chin, W. Liu, F. Théberge, Q. Luo, S. A. Hosseini, V. P. Kandidov, O. G. Kosareva, N. Aközbek, A. Becker, and H. Schroeder, "Some Fundamental Concepts of Femtosecond Laser Filamentation," in *Progress in Ultrafast Intense Laser Science III* (Springer Berlin Heidelberg, 2008), pp. 243-264.
15. J. Durnin, J. J. Miceli, and J. H. Eberly, "Diffraction-free beams," Phys. Rev. Lett. **58**, 1499-1501 (1987).
16. J. Durnin, "Exact solutions for nondiffracting beams. I. The scalar theory," J. Opt. Soc. Am. A **4**, 651-654 (1987).
17. F. O. Fahrbach, P. Simon, and A. Rohrbach, "Microscopy with self-reconstructing beams," Nat Photonics **4**, 780-785 (2010).
18. J. Arlt, K. Dholakia, L. Allen, and M. J. Padgett, "Efficiency of second-harmonic generation with Bessel beams," Physical Review A **60**, 2438-2441 (1999).
19. J. Durnin, J. H. Eberly, and J. J. Miceli, "Comparison of Bessel and Gaussian beams," Opt. Lett. **13**, 79-80 (1988).
20. Z. Song, Z. Zhang, and T. Nakajima, "Transverse-mode dependence of femtosecond filamentation," Optics express **17**, 12217-12229 (2009).
21. Z. Song and T. Nakajima, "Formation of filament and plasma channel by the Bessel incident beam in Ar gas: role of the outer part of the beam," Optics express **18**, 12923-12938 (2010).
22. M. Scheller, M. S. Mills, M. A. Miri, W. B. Cheng, J. V. Moloney, M. Kolesik, P. Polynkin, and D. N. Christodoulides, "Externally refuelled optical filaments," Nat Photonics **8**, 297-301 (2014).
23. J. H. Marburger, "Self-focusing: Theory," Progress in Quantum Electronics **4, Part 1**, 35-110 (1975).
24. J. Turunen, A. Vasara, and A. T. Friberg, "Holographic generation of diffraction-free beams," Applied Optics **27**, 3959-3962 (1988).
25. J. Strohaber, G. Kaya, N. Kaya, N. Hart, A. A. Kolomenskii, G. G. Paulus, and H. A. Schuessler, "In situ tomography of femtosecond optical beams with a holographic knife-edge," Opt. Express **19**, 14321-14334 (2011).
26. A. Brodeur and S. L. Chin, "Ultrafast white-light continuum generation and self-focusing in transparent condensed media," J. Opt. Soc. Am. B **16**, 637-650 (1999).
27. N. Kaya, J. Strohaber, A. A. Kolomenskii, G. Kaya, H. Schroeder, and H. A. Schuessler, "White-light generation using spatially-structured beams of

28. J. Strohaber, T. D. Scarborough, and C. J. G. J. Uiterwaal, "Ultrashort intense-field optical vortices produced with laser-etched mirrors," Applied Optics **46**, 8583-8590 (2007).
29. G. Steinmeyer and C. Bree, "Optical Physics: Extending filamentation," Nat Photonics **8**, 271-273 (2014).

femtosecond radiation," Optics express **20**, 13337-13346 (2012).